\documentstyle[12pt,psfig]{article}

\newlength{\titlesep}
\newlength{\authorsep}
\makeatletter

\def\fnum@figure{FIG.~\thefigure}

\newcounter{figureparent}
\@addtoreset{figure}{figureparent}

\@addtoreset{equation}{section}
\newcommand{\refeq}[1]{(\protect\ref{#1})}

\newcounter{eqnparent}
\@addtoreset{equation}{eqnparent}

\renewcommand{\abstract}{\if@twocolumn
  \section*{Abstract}
  \else
  \begin{center}
    {\bf Abstract\vspace{-.5em}\vspace{0pt}}
  \end{center}
  \quotation
  \fi}
\renewcommand{\endabstract}{\if@twocolumn\else\endquotation\fi}

\newcommand{\thismonth}{\ifcase\month\or
 January\or February\or March\or April\or May\or June\or
 July\or August\or September\or October\or November\or December\fi
 \space \number\year}

\newcommand{\preprintnumber}[1]
{\begin{flushright}
  \begin{tabular}{l} #1 \end{tabular}
  \end{flushright}}

\makeatother

\newcommand{\Rn}[1]{{\uppercase\expandafter{\romannumeral#1}}}
\newcommand{\gsim}%
{\mathrel{\mbox{\raisebox{-1.0ex}%
{$\stackrel{\textstyle >}{\textstyle \sim}$}}}}
\newcommand{\lsim}%
{\mathrel{\mbox{\raisebox{-1.0ex}%
{$\stackrel{\textstyle <}{\textstyle \sim}$}}}}

\newcommand{\kk}{$K^0$--$\overline{K}^0$ mixing}

\newcommand{\Journal}[4]{{#1} {\bf #2}, {#4} {(#3)}}
\newcommand{\pl}{\sl Phys.~Lett.}

\newcommand{\pr}{\sl Phys.~Rev.}

\newcommand{\prl}{\sl Phys.~Rev.~Lett.}
\newcommand{\np}{\sl Nucl.~Phys.}

\newcommand{\ptp}{\sl Prog.~Theor.~Phys.}

\newcommand{\zpc}{\sl Z.~Phys.~{\bf C}}

\newcommand{\epsfile}[1]{\relax}

\begin{document}
\baselineskip 18pt

\begin{titlepage}
\preprintnumber{%
KEK-TH-549 \\
KEK preprint 97-227 \\
}
\vspace*{\titlesep}
\begin{center}
{\LARGE\bf
Direct CP violation in  three-body decays of the $B$ meson
by resonance effects
}
\\
\vspace*{\titlesep}
{\large Ryoji Enomoto\footnote{enomoto@bsunsrv1.kek.jp}},\\
{\large Yasuhiro  Okada\footnote{okaday@theory.kek.jp}},\\
and {\large Yasuhiro  Shimizu\footnote{shimizuy@theory.kek.jp}}\\
\vspace*{\authorsep}
{\it KEK, Tsukuba, Ibaraki, 305-0801 Japan }
\\
\end{center}
\vspace*{\titlesep}
\begin{abstract}
We discuss direct CP violation in three-body decays  of $B$ meson
such as $B^\pm \to K^\pm\pi^+\pi^-$,
which involves various intermediate resonance states, such
as $B^\pm \to K^*_i \pi^\pm$ and $B^\pm \to \rho K^\pm$,
where $K^*_i$ represents the $\pi K$ resonance.
Due to the large final state interaction phases of the resonances,
the CP asymmetry can be as large as the 25\% level 
near the kinematical region where the $K^*_i$ and $\rho$ resonances
overlap. 
By examining a Dalitz plot of this mode and combining a measurement
of the branching ratio for $B^\pm \to \rho^\pm K$,
it is possible to extract the weak phase, 
$\phi_3$=arg$(-V_{ud}V_{ub}^*/(V_{cd}V_{cb}^*))$, of the
Kobayashi-Maskawa matrix.
\end{abstract}
\end{titlepage}
The origin of CP violation is one of the unresolved problems, although 
CP violation has been observed in the \kk\ system for 30 years.
CP violation occurs due to the complex phase of the Kobayashi-Maskawa
(KM) matrix \cite{KM} in the standard model (SM).
The main purpose of B-factory experiments is
to measure  the phases of the KM matrix, defined by
$\phi_1$=arg$(-V_{cd}V_{cb}^*/(V_{td}V_{tb}^*))$,
$\phi_2$=arg$(-V_{ud}V_{ub}^*/(V_{td}V_{tb}^*))$ and
$\phi_3$=arg$(-V_{ud}V_{ub}^*/(V_{cd}V_{cb}^*))$,
and to check the unitarity nature of the KM matrix \cite{belle,babar}.
It is known that $\phi_1$ can be measured in a so-called ``gold plated'' 
$B^0 \to J/\psi K_{s}$ mode \cite{BCS}. 
This mode is expected to determine $\phi_1$ to 
good accuracy. On the other hand,
measurements of $\phi_2$ and $\phi_3$ are considered to be rather
difficult from the experimental point of view.
Various methods to measure the weak phases are discussed in the
literature \cite{angle}.  
In order to observe the CP-violating effects it is necessary 
to have different CP-conserving and CP-violating phases in an amplitude.
The CP-violating phase is provided by the phase of the KM matrix.
For the neutral $B$ meson the CP-conserving phase is provided by the phase
of $B^0\overline{B}^0$ oscillation, whereas for the charged $B$ meson
the CP-conserving phase is dominantly supplied by the final-state-interaction
(FSI) phase. In general, it is difficult to reliably evaluate
the FSI phase because of hadron dynamics.  
However, we can reliably calculate the strong phase in the case of
resonances  by using the Breit-Wigner form. In addition, the resonance
provides a large  CP-conserving phase, which may lead to a large
direct CP-violation effect
\cite{AS} and the resonance enhancement of CP violation 
is discussed in several decay modes \cite{resonance,AEGS}.

In this paper we propose another method to measure the angle
$\phi_3$ using the direct CP violation in three-body decays of the $B$
meson, such as the $B^\pm \to K^\pm\pi^-\pi^+$ decay mode.
Several resonances contribute to this decay
mode, the effects which might enhance direct CP violation.
There are two channels 
of two-body intermediate states in this decay mode.
One is $B^\pm \to K^\pm \rho \to K^\pm\pi^-\pi^+$;
the other is $B^\pm \to K_i^* \pi^\pm \to K^\pm\pi^-\pi^+$, where 
$K_i^*$ denotes the $\pi K$ resonances summarized in Table.\ref{tab:Kres}
\cite{PDG}.
The $B^\pm \to K^\pm \rho$ process involves equally tree $(T)$ and QCD
penguin $(P)$ diagrams, whereas $B^\pm \to K^*_i \pi^\pm$ has a dominant
contribution from the QCD penguin $(P')$ diagram.\footnote{In principle
  there may be a tree-diagram contribution to  
$B^\pm \to K^*_i \pi^\pm$ through a rescattering process,
$B^\pm \to K^\pm \rho\to K_i^* \pi^\pm $
and this contribution will be discussed later.}
The amplitudes can be written by
\begin{eqnarray}
  \label{eq:bkpi}
  a_i &\equiv& A(B^- \to \overline{K}^*_i \pi^-)
  =  P_i'\ e^{i\delta_{P_i}'},
\\
  \label{eq:bkrho}
  b_\rho &\equiv& A(B^- \to K^- \rho)
  = T e^{i\delta_{T}} e^{- i\phi_3} + P\ e^{i\delta_{P}},
\end{eqnarray}
where $\delta_{T}$, $\delta_{P}$ and $\delta_{P_i}'$ represent
the strong phases, and the tree-diagram contribution has the weak
phase $\phi_3$. Concerning the corresponding amplitude for 
charge-conjugated modes $\overline{a_i}$ and $\overline{b_\rho}$,
only the sign of  $\phi_3$ should be changed. 
These amplitudes are normalized as
\begin{eqnarray}
  {\rm BR}(B^- \to \overline{K}^*_i\pi^-) = |a_i|^2,~~~
    {\rm BR}(B^- \to K^- \rho) = |b_\rho|^2.
\end{eqnarray}
Although these modes have not been observed experimentally,
the CLEO collaboration reported the branching ratios for 
the $B\to\pi K$ modes as follows \cite{CLEO}:
\begin{eqnarray}
  \label{eq:cleo}
  {\rm BR}(B^+ \to K^0\pi^+)
  = (2.3^{+1.1}_{-1.0}\pm 0.2 \pm 0.2) \times 10^{-5},
  \\
  {\rm BR}(B^0 \to K^+\pi^-)
  = (1.5^{+0.5}_{-0.4}\pm 0.1 \pm 0.1) \times 10^{-5}.
\end{eqnarray}
This result indicates that the $T$, $P$, $P'$ amplitudes for 
the $B^+ \to K^0\pi^+$ and $B^0 \to K^+\pi^-$ processes are the same
orders of magnitudes \cite{FM}:
Since the $B^0 \to\pi^+ K^-$ and $B^- \to \pi^- \overline{K}$ transitions
are essentially the same as the $B^- \to K^- \rho$ and
$B^- \to \overline{K}^*_i\pi^-$ transitions, respectively,
these intermediate contributions can significantly interfere with each
other in the $B^\pm \to K^\pm \pi^+ \pi^-$ decay modes.
The intermediate resonances, which subsequently decay to two
pseudo-scalars, have large FSI phases.
We assume that these resonance contributions can be treated by the
Breit-Wigner form:
\begin{eqnarray}
  \Pi_{K_i^*}(s) = 
  \frac{\sqrt{\pi\Gamma_{K^*_i}/m_{K^*_i}}}
  {s-m_{K^*_i}^2+im_{K^*_i}\Gamma_{K^*_i}},~~
  \Pi_{\rho}(s') = 
  \frac{\sqrt{\pi\Gamma_{\rho}/m_{\rho}}}
    {s'-m_{\rho}^2+im_{\rho}\Gamma_{\rho}},
\end{eqnarray}
where $s=(p_{K^-}+p_{\pi^+})^2$, $s'=(p_{\pi^+} + p_{\pi^-})^2$
and the $m_i$'s and $\Gamma_i$'s are the masses and widths of the
resonances, respectively. 

Assuming that the $B^- \to K^-\pi^-\pi^+$ decay amplitude is dominated
by the above resonances, we can write the decay amplitude in a
narrow-width approximation as follows:
\begin{eqnarray}
  \label{eq:amp}
  A(B^- \to K^-\pi^-\pi^+)
  &=&
  \sum_{i}
  \frac{a_i}
  {m_{K^*_i}\sqrt{\beta_{K^*_i}^{B\pi}}}
  \sqrt{\frac{2{\rm B}_{K^*_i}^{K\pi}}
    {\beta_{K^*_i}^{K\pi}}}
  \ \Pi_{K_i^*}(s)\ \Theta(z)
  \nonumber \\
  &+&
  \frac{b_\rho}
  {m_{\rho}\sqrt{\beta_{\rho}^{BK}}}
  \sqrt{\frac{2{\rm B}_{\rho}^{\pi \pi}}
    {\beta_{\rho}^{\pi\pi}}}
  \ \Pi_{\rho}(s')\ \Theta(z'),
\end{eqnarray}
where ${\rm B}_{\rho}^{\pi \pi}$ and ${\rm B}_{K^*_i}^{K \pi}$ 
denote the branching ratios for $\rho \to \pi\pi$ and
$K^*_i \to K \pi$, respectively, and $\beta_i^{jk}$ is defined by
\begin{eqnarray}
  \label{eq:beta}
  \beta_i^{jk}
  = \sqrt{(1-(m_j+m_k)^2/m_i^2)(1-(m_j-m_k)^2/m_i^2)}~~.
\end{eqnarray}
The function $\Theta$ describes the angular component of the
amplitude for the resonance decay \cite{AEGS,LNQS}.
The explicit form  depends on the spin
of the resonance, and is given by 
\begin{eqnarray}
  \label{eq:theta}
  \Theta(z) = 
  \left\{
      \begin{array}{lc}
        \sqrt{\frac{1}{2}} ,        & ~~\mbox{spin 0},\\
        \sqrt{\frac{3}{2}}z,        & ~~\mbox{spin 1},\\
        \sqrt{\frac{5}{8}}(3z^2-1), & ~~\mbox{spin 2},
      \end{array}
    \right.
\end{eqnarray}
where $z(z')$ denotes $\cos\theta (\cos\theta')$, 
and $\theta(\theta')$ is the angle between the momentum of the $B$
and $K^-(\pi^-)$ mesons in the center-of-mass frame of the 
$\pi^+\pi^-$ $(\pi^+ K^-)$ pairs.
$z(z')$ can be written as a function of $s$, $s'$ as follows:
\begin{eqnarray}
  \label{eq:angle}
  z &=& \frac{
    2ss'+s^2-s(m_B^2+m_K^2+2m_\pi^2)
    +(m_B^2-m_\pi^2)(m_K^2-m_\pi^2)
    }{
    \sqrt{
      (s-(m_K^2+m_\pi)^2)(s-(m_K-m_\pi)^2)(s-(m_B+m_\pi)^2)(s-(m_B-m_\pi)^2)
      }
    },~~~~~
\\
  z' &=& \frac{
    \sqrt{s'}(2s + s' -m_B^2 - m_K^2 - 2 m_\pi^2)
    }{
    \sqrt{
      (s'-4 m_\pi^2)(s'-(m_B+m_K)^2)(s'-(m_B-m_K)^2)
      }
    }.
\end{eqnarray}

\begin{figure}[htbp]
  \begin{center}
    \leavevmode
    \centerline{\psfig{file=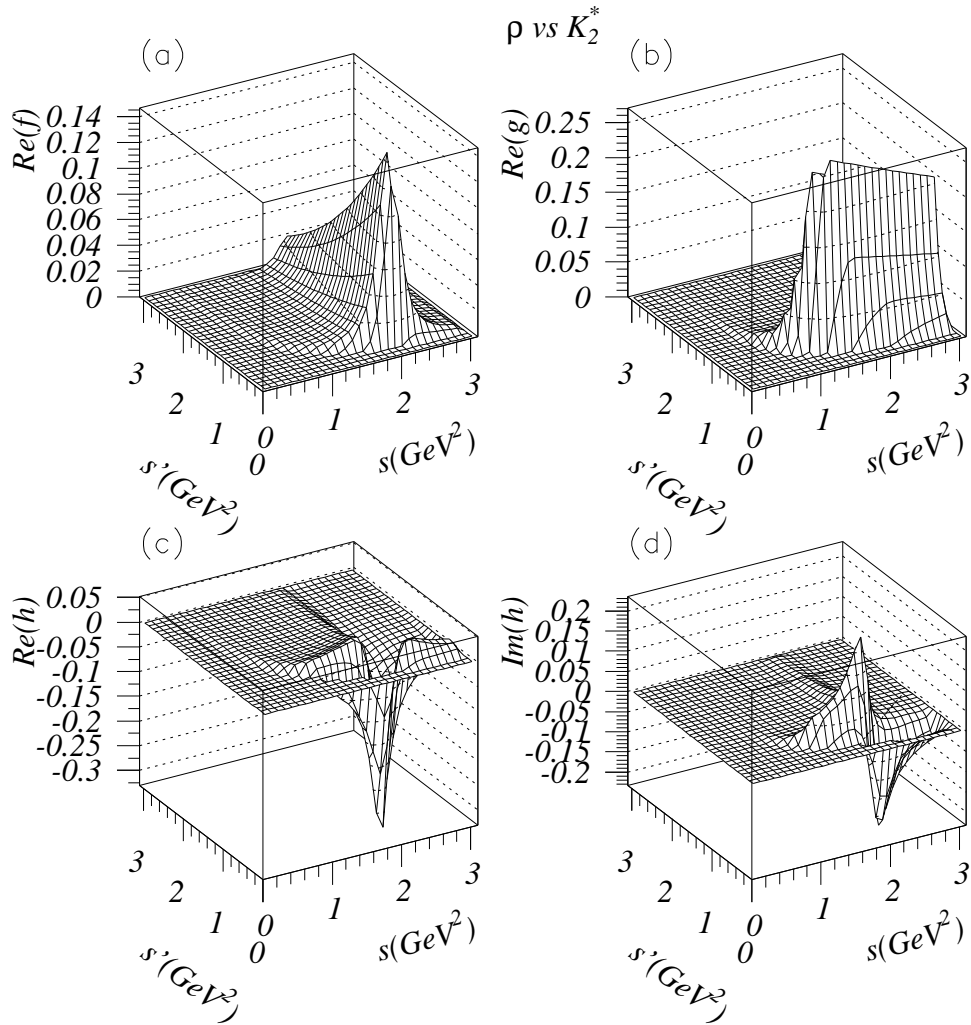}}    
    \caption{Kinematical functions defined in Eq.\refeq{eq:func}.
      We take into account the $\rho$ and $K^*_2(1430)$ resonances
      for the $\pi\pi$ and $K\pi$ channels, respectively.
      $s=(p_{K^-}+p_{\pi^+})^2$ and $s'=(p_{\pi^-}+p_{\pi^+})^2$. }
    \label{fig:1}
  \end{center}
\end{figure}

For simplicity, we hereafter consider the case that only the $K_2^*(1430)$
and $\rho$  resonances contribute to each $\pi K$ and $\pi \pi$
channel, respectively.\footnote{The $K_2^*(1430)$ resonance overlaps with
the $K_0^*(1430)$ resonance but we neglect it here.
We will discuss this contribution later.}
We later discuss more general formula which includes
all of the intermediate resonances.
The differential decay width for $B^- \to K^-\pi^-\pi^+$
can be written as
\begin{eqnarray}
  \label{eq:width}
  d\Gamma(B^- &\to& K^-\pi^-\pi^+)/dsds'
  \\ \nonumber
  &=&
  \left|a_{K^*_2}\right|^2
  \mbox{Re}\left(f^{(K^*_2,K^*_2)}(s,s')\right)
  +
  \left|b_{\rho}\right|^2
  \mbox{Re}\left(g^{(\rho,\rho)}(s,s')\right)
  \\ \nonumber 
  &+&
    \mbox{Re}\left(a_{K^*_2}\ b_{\rho}^* \right)
    \mbox{Re}\left(h^{(K^*_2,\rho)}(s,s')\right)
    -\mbox{Im}\left(a_{K^*_2}\ b_{\rho}^* \right)
    \mbox{Im}\left(h^{(K^*_2,\rho)}(s,s')\right).
\end{eqnarray}
The distribution on a Dalitz plot is described by the functions 
$f$, $g$ and $h$ in this formula, and are written as
\begin{eqnarray}
  \label{eq:func}
  f^{(K_i^*,K_j^*)}(s,s') 
  &=&
  \frac{2}{m_{K_i^*}m_{K_j^*}}
  \sqrt{
    \frac{{\rm B}_{K_i^*}^{K\pi}{\rm B}_{K_j^*}^{K\pi}}
    {\beta_{K_i^*}^{B\pi}\beta_{K_j^*}^{B\pi}
     \beta_{K_i^*}^{K\pi}\beta_{K_j^*}^{K\pi}}}
   \left(
   \Pi_{K^*_i}(s)\Pi_{K^*_j}(s)^*
   \right)
   \Theta_{K_i^*}^{\pi K}(z)\Theta_{K_j^*}^{\pi K}(z),~~~~~~~
\\
  g^{(f_i,f_j)}(s,s') 
  &=&
  \frac{2}{m_{f_i}m_{f_j}}
  \sqrt{
    \frac{{\rm B}_{f_i}^{\pi\pi}{\rm B}_{f_j}^{\pi\pi}}
    {\beta_{f_i}^{BK}\beta_{f_j}^{BK}
     \beta_{f_i}^{\pi\pi}\beta_{f_j}^{\pi\pi}}}
 \left(
   \Pi_{f_i}(s')\Pi_{f_j}(s')^*
   \right)
   \Theta_{f_i}^{\pi \pi}(z')\Theta_{f_j}^{\pi \pi}(z'),~~~~~~~
\\
  h^{(K_i^*,f_j)}(s,s') 
  &=&
  \frac{4}{m_{K^*_i}m_{f_j}}
  \sqrt{
    \frac{{\rm B}_{K^*_i}^{K\pi}{\rm B}_{f_j}^{\pi\pi}}
    {\beta_{K^*_i}^{B\pi}\beta_{f_j}^{BK}
     \beta_{K^*_i}^{K\pi}\beta_{f_j}^{\pi\pi}}}
 \left(
   \Pi_{K^*_i}(s)\Pi_{f_j}(s')^*
   \right)
   \Theta_{K^*_i}^{K \pi}(z)\Theta_{f_j}^{\pi \pi}(z').
\end{eqnarray}
The above expressions are defined for general meson states,
$f_i$=$\rho$, $f_0$, or $f_2$, which decay into the 2$\pi$ state
since we will later discuss the $B^\pm \to f_i \pi^\pm \to \pi^+\pi^-K^\pm$
mode in addition to the $B^\pm \to \rho K^\pm \to \pi^+\pi^-K^\pm$ mode.
We can obtain the differential width for the charge-conjugated mode
by simply changing the amplitude $a_i(b_\rho)$ into
$\overline{a}_i(\overline{b}_\rho)$.
In Fig.\ref{fig:1}(a)-(d) we show the kinematical functions
Re$(f^{(K^*_2,K^*_2)})$, Re$(g^{(\rho,\rho)})$, Re$(h^{(K^*_2,\rho)})$ and 
Im$(h^{(K^*_2,\rho)})$ in the region $R=\{0<s,s'<3 \mbox{GeV}^2\}$.
It can be seen that the signs of Re$(f^{(K^*_2,K^*_2)})$ and 
Re$(g^{(\rho,\rho)})$ are always positive, whereas Re$(h^{(K^*_2,\rho)})$ and 
Im$(h^{(K^*_2,\rho)})$  change their signs around the 
$K^*_2(1430)$ and $\rho$ poles.
Note that the $K^*_2(1430)$ and $\rho$ 
contributions can overlap around the region $s\simeq m_{K^*_2}^2$,
$s'\simeq m_{\rho}^2$,  and that there is a large interference in this
region. 

Let us discuss how we can determine the parameters in Eq.\refeq{eq:width}. 
Since one of the strong phases is irrelevant, we have six unknown 
parameters ($T$, $P$, $P'$, $\delta_T - \delta_P'$, $\delta_P-\delta_P'$ 
and $\phi_3$) in this formula.
We can determine these parameters as follows.
In the kinematical region $s \simeq m_{K^*_2}^2$ and
$s' \gg m_{\rho}^2$, the $K^*_2$ contribution dominates in the decay
width. We can extract the amplitudes $|a_{K^*_2}|$ and 
$|\overline{a}_{K^*_2}|$ by retaining only the first term in the
Eq.\refeq{eq:width} and the parameter $P'$ can be determined as 
$P'=|a_{K^*_2}|$. By examining the distribution on the Dalitz plot in the
overlapping region $s \simeq m_{K^*_2}^2$ and $s' \simeq m_{\rho}^2$, 
we can extract Re$(a_{K^*_2}b_{\rho}^*)$, Im$(a_{K^*_2}b_{\rho}^*)$,
Re$(\overline{a}_{K^*_2}\overline{b}_{\rho}^*)$ and
Im$(\overline{a}_{K^*_2}\overline{b}_{\rho}^*)$.
Since we have already known $|a_{K^*_2}|$ and  $|\overline{a}_{K^*_2}|$,
we can determine Re$(b_\rho e^{i\delta_P'})$, 
Im$(b_\rho e^{i\delta_P'})$, Re$(\overline{b}_\rho e^{i\delta_P'})$
and Im$(\overline{b}_\rho e^{i\delta_P'})$.
There are five unknown 
parameters in $b$ and $\overline{b}$ but we have only four constraints.
If we combine the branching ratio for $B^- \to \rho^- \overline{K}$,
whose amplitude is written as 
\begin{eqnarray}
  \label{eq:isospin}
  A(B^- \to \rho^- \overline{K}) 
  =  -\sqrt{2} P e^{i\delta_P},
\end{eqnarray}
by using the isospin symmetry \cite{DGR},
we can therefore fix all the parameters, including the weak phase,
$\phi_3$.
Note that we assume that there is no tree-diagram contribution to 
the $B^- \to \overline{K}^*\pi^-$ amplitude in Eq.\refeq{eq:bkpi}.
In general the tree-diagram contribution can be induced through the 
rescattering process such as $B^-\to \rho K^- \to \overline{K}^*_2\pi^-$.
If we take into account the tree-diagram contribution, which can be
written as $T'e^{i\delta_{T}'}e^{-i\phi_3}$,
we have two unknown parameters, $\delta_T'$ and $T'$, in addition to
the above six unknown parameters.
We cannot determine $\phi_3$ since there are only seven independent 
constraints even if we combine the branching ratio for 
$B^\pm \to \rho^\pm \overline{K}$.
Although it is hard to calculate the corrections from the rescattering 
effects quantitatively, it is estimated to be of order 10 \% 
for the $B \to \pi K$ modes \cite{FSI}.
Unless this rescattering effect dominates in the amplitudes,
we can expect to determine $\phi_3$ with reasonable accuracy.

The CP-violating quantity ${\cal A}_{CP}$ is defined as
\begin{eqnarray}
  \label{eq:acp}
  {\cal A}_{CP} \equiv 
  \frac{d\Gamma(B^- \to K^-\pi^+\pi^-)/dsds'
  - d\Gamma(B^+ \to K^+\pi^+\pi^-))/dsds'}
  {d\Gamma(B^- \to K^-\pi^+\pi^-)/dsds'
  + d\Gamma(B^+ \to K^+\pi^+\pi^-))/dsds'}~~,
\end{eqnarray}
where we can write the denominator and numerator as follows:
\begin{eqnarray}
  \label{eq:deno}
  d\Gamma(B^- &\to& K^-\pi^+\pi^-)/dsds'
  + d\Gamma(B^+ \to K^+\pi^+\pi^-)/dsds'
\nonumber\\
&=&
  2P'^2 
  \mbox{Re}\left(f^{(K^*_2,K^*_2)}\right)
  +2\left(T^2+P^2+2TP\cos\delta_P\cos\phi_3 \right) 
  \mbox{Re}\left(g^{(\rho,\rho)}\right)
\nonumber\\
  &+& 2P'\left( T \cos\delta_P'\cos\phi_3 +
    P\cos(\delta_P'-\delta_P)\right) \mbox{Re}\left(h^{(K^*_2,\rho)}\right)
\nonumber\\
  &-& 2P'\left( T \sin\delta_P'\cos\phi_3 +
    P\sin(\delta_P'-\delta_P)\right)
  \mbox{Im}\left(h^{(K^*_2,\rho)}\right),
\\
  \label{eq:num}
  d\Gamma(B^- &\to& K^-\pi^+\pi^-)/dsds'
  - d\Gamma(B^+ \to K^+\pi^+\pi^-)/dsds'
\nonumber\\
&=&
  -2T\left[2P\sin\delta_P\mbox{Re}\left(g^{(\rho,\rho)}\right)
  +P'\sin\delta_P'\mbox{Re}\left(h^{(K^*_2,\rho)}\right)
\right.
\nonumber\\
  &&+ 
\left.
 P'\cos\delta_P'
    \mbox{Im}\left(h^{(K^*_2,\rho)}\right)\
  \right]\sin\phi_3.
\end{eqnarray}
The CP asymmetry appears as an asymmetry of the distribution on the
Dalitz plot for $B^+$ and $B^-$ decays. Let us consider here the 
integrated asymmetry in the region $R$. When we integrate the functions
Re$(h^{(K^*_2,\rho)})$ and Im$(h^{(K^*_2,\rho)})$ in this region,
they almost vanish because the functions change their signs
around the resonance poles,
\begin{eqnarray}
  \label{eq:deno2}
  &\int_R&\left(d\Gamma(B^- \to K^-\pi^+\pi^-)/dsds'
  + d\Gamma(B^+ \to K^+\pi^+\pi^-)/dsds'\right)
\nonumber\\
&\simeq&
\int_R\left[
  2P'^2 \mbox{Re}\left(f^{(K^*_2,K^*_2)}\right)
  +2\left(T^2+P^2+2TP\cos\delta_P\cos\phi_3 \right) 
  \mbox{Re}\left(g^{(\rho,\rho)}\right)
\right].
\end{eqnarray}
In order to see how the CP asymmetry can be large,
it is necessary to evaluate the hadronic matrix elements, such as
$T$, $P$ and $P'$, which we cannot calculate reliably.
Instead of evaluating the hadronic matrix elements based on a
specific hadronic model \cite{hadron}, let us make a crude estimation 
assuming the relation $T=P=P'$.
We can obtain the maximal CP asymmetry by choosing the integration of
Eq.\refeq{eq:num}. If we define 
\begin{eqnarray}
  \label{eq:acp2}
  {\cal A}_R \equiv
  \frac{\int_R ds ds'
    \left|\mbox{Re} \left(
        h^{(K^*_2,\rho)}(s,s')\right)\right|}
  {\int_R ds ds'\left[\mbox{Re}\left(f^{(K^*_2,K^*_2)}(s,s')\right) +
    2\mbox{Re}\left(g^{(\rho,\rho)}(s,s')\right)\right]},
\\
  \label{eq:acp3}
  {\cal A}_I \equiv
  \frac{\int_R ds ds'
    \left|\mbox{Im} \left(h^{(K^*_2,\rho)}(s,s')\right)\right|}
  {\int_R ds ds'\left[\mbox{Re}\left(f^{(K^*_2,K^*_2)}(s,s')\right) +
    2\mbox{Re}\left(g^{(\rho,\rho)}(s,s')\right)\right]},
\end{eqnarray}
${\cal A}_R$ and ${\cal A}_I$ correspond to the
optimal CP asymmetry in the case of $\delta_P=0$,
$\delta_P'=\phi_3=\pi/2$  and  $\delta_P=\delta_P'=0,\phi_3=\pi/2$,
respectively.\footnote{If $\delta_P$ is large enough,
it is possible to have a large direct CP violation in $B^\pm \to K^\pm
\rho$. Hence we may observe the direct CP violation even in the region
$s\gg m_{K^*_2}^2$, $s' \simeq m_{\rho}^2$.}
We also calculate CP asymmetry using other $\pi\pi$ resonances,
such as $f_0$ and $f_2$ summarized in Table.\ref{tab:pires}.
Although the isospin relation cannot be used for $f_0$, $f_2$
and therefore we cannot determine $\phi_3$ from these modes,
a large direct CP asymmetry may be observable.
In Table.\ref{tab:asymmetry} we present the values of ${\cal A}_R$ and
${\cal A}_I$ not only in case of $K^*_2$ and $\rho$,  but also in case of
other resonances. 
Although the actual CP asymmetry indeed depends on the poorly known 
value such as $\phi_3$, $\delta_P$ and $\delta_P'$,
it is reasonable to expect that the CP asymmetry can be as large as
25\% level at somewhere around the overlapping region.

It is straitforward to write a general formula which includes
all of the intermediate resonances as follows:
\begin{eqnarray}
  \label{eq:width2}
  d\Gamma(B^- &\to& K^-\pi^-\pi^+)/dsds'
  \\ \nonumber
  &=&
  \sum_{i}\left|a_i\right|^2
  \mbox{Re}\left(f^{(K_{i}^*,K_{i}^*)}(s,s')\right)
  +
  \sum_{i}\left|b_i\right|^2
  \mbox{Re}\left(g^{(f_{i},f_{i})}(s,s')\right)
  \\ \nonumber 
  &+&
  2\sum_{i<j}
  \left\{
    \mbox{Re}\left( a_i\ a_j^*\right)
    \mbox{Re}\left(f^{(K_i^*,K_j^*)}(s,s') \right)
  - \mbox{Im}\left(a_i\ a_j^*\right)
    \mbox{Im}\left(f^{(K_i^*,K_j^*)}(s,s')\right)
  \right\}
  \\ \nonumber 
  &+&
  2\sum_{i<j}
  \left\{
    \mbox{Re}\left(b_i \ b_j^* \right)
    \mbox{Re}\left(g^{(f_i,f_j)}(s,s') \right)
  - \mbox{Im}\left(b_i \ b_j^* \right)
    \mbox{Im}\left(g^{(f_i,f_j)}(s,s')\right)
  \right\}
  \\ \nonumber 
  &+&
  \sum_{i,j}
  \left\{
    \mbox{Re}\left(a_i\ b_j^* \right)
    \mbox{Re}\left(h^{(K_i^*,f_j)}(s,s') \right)
  - \mbox{Im}\left(a_i\ b_j^* \right)
    \mbox{Im}\left(h^{(K_i^*,f_j)}(s,s')\right)
  \right\}.
\end{eqnarray}
Since the mass of $K^*_0(1430)$ is as same as that of $K^*_2(1430)$,
these resonances coincide at the kinematical region $\sqrt{s} 
\simeq 1430$ MeV. Even if the masses are degenerate, we can determine 
$|a_{K^*_0}|$, $|a_{K^*_2}|$, Re$(a_{K^*_0}a_{K^*_2})$ and 
Im$(a_{K^*_0}a_{K^*_2})$ by analyzing the Dalitz plot in the region 
$s \simeq m_{K_0^2}$ and $s' \gg m_{\rho}^2$
since the $K^*_0$ and $K^*_2$ contributions depend differently on $s'$.
Once $a_{K^*_0}$ and $a_{K^*_2}$ are known up to an overall phase,
we can extract Re$(b_j)$, Im$(b_j)$, Re$(\overline{b}_j)$ and 
Im$(\overline{b}_j)$
similarly as in the case of $K^*_2$ alone by examining the Dalitz plot
in the overlapping region. 
In case of the $\rho$ resonance we can again obtain another constraint
from $B^- \to \rho^- \overline{K}$ with the isospin symmetry.
We can therefore determine all the parameters including $\phi_3$.
In case of $f_0$ and $f_2$ resonances we cannot determine $\phi_3$ 
although a large direct CP violation may be observable.

Here, we discuss  the experimental feasibility to detect this CP 
violation. The detector and accelerator are assumed to be 
similar to the Belle detector
in the KEK B-factory \cite{belle}. The decays of $B\rightarrow 
K^*_2(1430)\pi$ and $\rho K$ are used. Both  branching 
fractions, i.e., $\mbox{BR}(B\rightarrow K^*_2\pi)\cdot 
\mbox{BR}(K^*_2\rightarrow K^+\pi^-)$ and $\mbox{BR}(B\rightarrow \rho K)\cdot
\mbox{BR}(\rho \rightarrow \pi^+\pi^-)$, are set to $10^{-5}$.
The kinematical region of $s<3\mbox{GeV}^2$ and $s'<3\mbox{GeV}^2$ is
about 12\% of the total phase space, considering the experimental
acceptance and decay angular distributions. The total experimental
acceptance is about 3\%, including the geometrical factor and
other cuts, such as to suppress continuum
background ($e^+e^-\to (\gamma) \to q\bar{q}$).
The signal-to-noise ratio is estimated to be $\sim$2.
In the case of having a 25\% direct CP asymmetry, 
$\sim 300\ \mbox{fb}^{-1}$ of the data on $\Upsilon(4s)$ is necessary to
obtain a $3\sigma$ asymmetry. 
A significant improvement can be expected, if the kinematical region is
reduced to only around the resonance overlapping region.

In this paper we consider the direct CP violation in three-body decays,
such as $B^\pm \to K^\pm\pi^+\pi^-$.
The $B$ meson can decay to the three-body final state through 
two different channels, $K^*_i \pi^\pm$ and $\rho K^\pm$, and
these amplitudes may interfere with the same orders of magnitude 
near the kinematical region where the $K^*_i \pi^\pm$
and $\rho K^\pm$ contributions overlap.
The resonances $K^*_i$ and $\rho$ 
provide a large final-state-interaction phase,
which is reliably calculated using the Breit-Wigner form.
This yields a large CP asymmetry around the overlapping region,
and we can expect the CP asymmetry to be
as large as the 25\% level in this region.
By measuring the distribution on the Dalitz
plot and combining the measurement of branching ratio for 
$B^\pm \to \rho^\pm K$, it is possible to extract the weak phase,
$\phi_3$. 

This work was supported in part by the
Grant-in-Aid of the Ministry of Education, Science, Sports and
Culture, Government of Japan.

\newpage

%
\begin{table}[htbp]
  \begin{center}
    \leavevmode
    \begin{tabular}{|c||c|c|c|c|}
      \hline
      \hline
      $K_i^*$ & spin & mass (MeV) & width (MeV)
      & {\rm B}$(\overline{K}_i^* \to K^- \pi^+)$\\
      \hline
      \hline
      $K^*$(892) & 1 &  896 & 51 & 0.67 \\
      \hline
      $K^*_0$(1430) & 0 & 1429 & 287 & 0.62 \\
      \hline
      $K^*_2$(1430) & 2 & 1432 & 109 & 0.33 \\
      \hline
      \hline
    \end{tabular}
    \caption{Summary of $K\pi$ resonances.}
    \label{tab:Kres}
  \end{center}
\end{table}
\begin{table}[htbp]
  \begin{center}
    \leavevmode
    \begin{tabular}{|c||c|c|c|c|}
      \hline
      \hline
      $f_i$ & spin & mass (MeV) & width (MeV) 
      & {\rm B}$(f_i \to \pi^+ \pi^-)$\\
      \hline
      \hline
      $\rho$(770) & 1 &  769 & 151 & 1 \\
      \hline
      $f_0$(980) & 0 & 980 & 40-100 & 0.52 \\
      \hline
      $f_2$(1270) & 2 & 1275 & 185 & 0.57 \\
      \hline
      \hline
    \end{tabular}
    \caption{Summary of $\pi\pi$ resonances.}
    \label{tab:pires}
  \end{center}
\end{table}
\begin{table}[htbp]
  \begin{center}
    \leavevmode
    \begin{tabular}{|c|c||c|c|}
      \hline
      \hline
      $K_i^*$ \,& $f_i$ \,& ${\cal A}_R$ & ${\cal A}_I$ \\
      \hline
      \hline
      $K^*$ \, & $\rho$ \,& 0.13 & 0.087 \\
      \hline
      $K^*$ \, & $f_0$ \, & 0.15 & 0.10 \\
      \hline
      $K^*$ \, & $f_2$ \, & 0.26  & 0.25  \\
      \hline
      $K_0$ \, & $\rho$ \,& 0.20  & 0.22  \\
      \hline
      $K_0$ \, & $f_0$  \,& 0.26  & 0.28  \\
      \hline
      $K_0$ \, & $f_2$ \, & 0.26 & 0.27  \\
      \hline
      $K_2$ \, & $\rho$ \, & 0.26 & 0.25  \\
      \hline
      $K_2$ \, & $f_0$ \, & 0.29 & 0.26  \\
      \hline
      $K_2$ \, & $f_2$ \, & 0.26 & 0.25  \\
      \hline
    \end{tabular}
    \caption{CP asymmetries defined in Eq.\refeq{eq:acp2}-\refeq{eq:acp3}.
      Each row corresponds to the asymmetries,
      taking into account  only one resonance for
      each $\pi\pi$ and $\pi K$ channel.
      Here the width for $f_0$ is assumed to be 70 MeV.}
    \label{tab:asymmetry}
  \end{center}
\end{table}

\end{document}